\documentclass[]{spie}

\usepackage{graphicx,color}

\title{Wide field imaging for the Square Kilometre Array}
\author{T.J. Cornwell\supit{a}, M. A. Voronkov\supit{a} and B. Humphreys\supit{a}
\skiplinehalf
\supit{a}CSIRO Astronomy and Space Science, CSIRO, PO Box 76, Epping, New South Wales, Australia \\
}

\authorinfo{Further author information: Send correspondence to T.J.C.: E-mail: tim.cornwell@csiro.au, Telephone: +61 2 9372 4261}

\begin{document}
\maketitle
\begin{abstract}

Wide-field radio interferometric telescopes such as the Square Kilometre Array now being designed are subject to a number of aberrations. One particularly pernicious aberration is that due to non-coplanar baselines whereby long baselines incur a quadratic image-plane phase error. There are numerous algorithms for dealing with the non-coplanar baselines effect. As a result of our experience with developing processing software for the Australian Square Kilometre Array Pathfinder, we advocate the use of a hybrid algorithm, called $w$ snapshots, based on a combination of $w$ projection and snapshot imaging. This hybrid overcomes some of the deficiencies of each and has advantages from both. Compared to pure $w$ projection, $w$ snapshots uses less memory and execution time, and compared to pure snapshot imaging, $w$ snapshots uses less memory and is more accurate. At the asymptotes, $w$ snapshots devolves to $w$ projection and to snapshots.

\end{abstract}


\section{THE SQUARE KILOMETRE ARRAY}
\label{sec:thesquarekilometrearray}

The Square Kilometre Array (SKA)\cite{Dewdney2011} is a world class radio telescope now being constructed by an international consortium of close to a dozen countries. It is being constructed in two phases, Phase 1 planned to start observations in 2018, and a subsequent roughly ten times larger Phase 2 planned to start observations in 2022. SKA1 covers observing frequencies from 70MHz to 3GHz, with three distinct receptor technologies.
\begin{itemize}
\item SKA1\_DISH An array of 250 15m diameter parabolic dishes with single pixel feeds,
\item SKA1\_SURVEY An array of 96 15m diameter parabolic dishes with phased array feeds\cite{}
\item SKA1\_AA\_LOW An array of 250 aperture array stations, each station having 11,400 active antennas (roughly dipoles).
\end{itemize}

In Phase 2, SKA1\_AA\_LOW would be grown to lower baselines, SKA1\_DISH would be grown in number of antennas by close to an order of magnitude, and depending on technology demonstration, there may also be a mid-frequency range aperture array telescope.

Taken together, these telescopes will form the Square Kilometre Array Observatory. In Phase 1, SKAO will be located in both Australia (SKA1\_SURVEY, SKA1\_AA\_LOW) and South Africa (SKA1\_DISH). 

The science case for SKA is very well-developed\cite{Carilli2004}. It spans a wide range of pressing topics in astronomy and astrophysics.

The data processing for the SKA will be very demanding. High sensitivity implies many measurements for processing. In addition, high sensitivity imaging requires high dynamic range since at these wavelengths the radio sky is bright with sources. In addition, all three telescopes will have wide fields of view, for which the processing is intrinsically demanding. It is this last aspect we concentrate on in this paper.  Over the last 5 years, we have been engaged in developing the Australia Square Kilometre Array Pathfinder (ASKAP) \cite{2009IEEEP..97.1507D}, which is a smaller version of SKA1\_SURVEY. As part of the construction of ASKAP, we have developed a software system, ASKAPSoft, for processing ASKAP observations into science products. We will describe what we believe to be the optimal wide-field processing algorithm for ASKAP data and how it will scale to SKA.

\section{WIDE FIELD IMAGING}
\label{sec:widefieldimaging}

A major obstacle to wide field imaging  with radio synthesis arrays is the non-coplanar baselines effect\cite{CornwellPerley1992}. This prevents the use of a simple two-dimensional transform to form an image of the sky from the measured visibilities.

To understand the $w$ term, we first recall that for small fields of view,  the visibility obeys: 
\begin{equation}
\label{eqn:me2d}
V(u,v)=\int\nolimits I(l,m)e^{j2\pi\left(ul+vm\right)}\;dldm\mbox{.}
\end{equation}
For larger fields of view, the $w$-term becomes important
\begin{equation}
\label{eqn:me3d}
V(u,v)=\int\nolimits\frac{I(l,m)}{\strut\sqrt{\strut 1-l^2-m^2}}\;e^{\;j2\pi\left(ul+vm+
w\left(\sqrt{\strut 1-l^2-m^2}-1\right)\right)}\;dldm\mbox{.}
\end{equation}
The new term $w$ is the component of antenna-antenna vector towards the phase centre of the field of view. The physical origin of this phase term is straightforward -- it comes from the need to refer the electric field to the same physical plane. This requires a Fresnel term. It is straightforward to show that the $w$-term effect is significant if the field of view is comparable to or greater than the square root of the resolution (both measured in radians):

\begin{equation}
\label{eqn:sigwterm}
\theta_{FOV}\ge\sqrt{\theta_{resolution}}
\end{equation}

The actual maximum allowed field of view depends upon the accuracy required in the peak fluxes. To track this, we introduce a parameter $\alpha_{FOV}$ that quantifies how much smaller the field of view must be to attain a given level of accuracy. We will return to the estimate of this parameter later.

\begin{equation}
\label{eqn:sigwterm1}
\theta_{FOV}\ge\alpha_{FOV} \sqrt{\theta_{resolution}}
\end{equation}

We can now discuss the various algorithms for correcting the $w$ term.  There are three approaches to counter the non-coplanar baselines effect in use in ASKAPsoft - wprojection \cite{2008ISTSP...2..647C}, wstacking, 
snapshots \cite{CornwellPerley1992,Ord:2010tt}, and 
faceting\cite{CornwellPerley1992}.

\begin{description}
\item[W Projection] The $w$ term can be expressed as a multiplicative effect in image space, or a convolution in Fourier space. In 
both cases, the $w$ variable acts as a parameter. The convolution relationship between visibility on $w=0$ and an arbitrary $w$ plane is:
\begin{equation}
\label{eqn:wprojrel}
\begin{array}{l}
V(u,v,w)=\tilde G(u,v,w)\otimes V(u,v)\\
\tilde G(u,v,w)=\int\nolimits\frac{\strut e^{\;j2\pi w(\sqrt{1-l^2-m^2}-1)}}
{\strut\sqrt{1-l^2-m^2}}\;e^{\;j2\pi\left(ul+vm\right)}\;dldm\\
\tilde G(u,v,w) \approx 
\frac{ e^{\;j\pi 
                \frac{u^2+v^2}{w}
             }}{jw}
\end{array}
\end{equation}

In figure \ref{fig:w3d}, we show the three dimensional form of $G(u,v,w)$, with the $w$ axis running vertically. The envelope of this function obeys $u/w \sim \Theta_{FOV}$, and the number of fringes across the function scales as $w^{\frac{3}{2}}$\cite{Humphreys2011}.

\begin{figure}
\caption{The three dimensional form of $G(u,v,w)$, with the $w$ axis running vertically. It is Hermitean symmetric about the centre point, shows in red here.}
\label{fig:w3d}
\begin{center}
\includegraphics[scale=1.0]{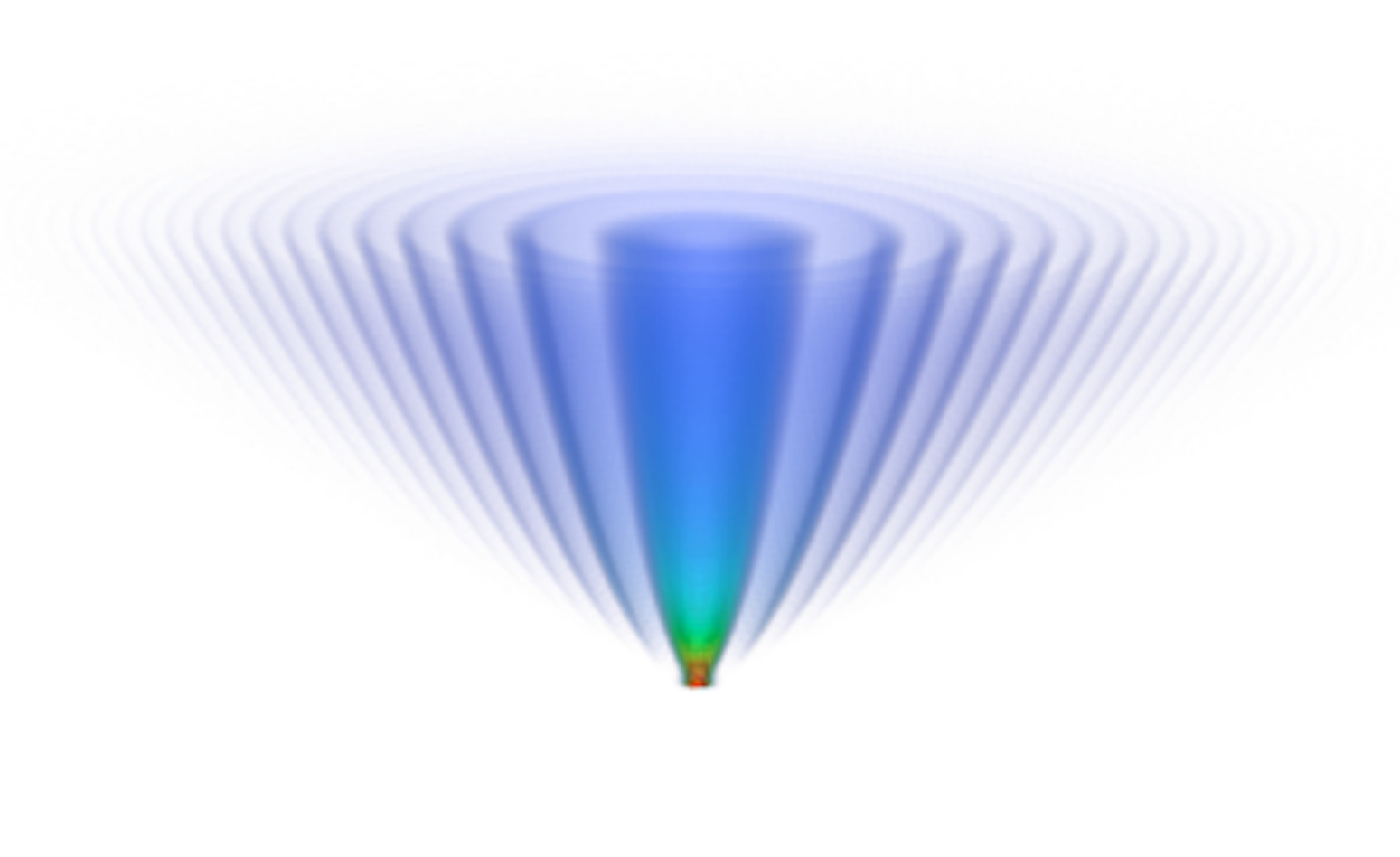}
\end{center}
\end{figure}

Before going further, we need to review processing for small fields where the $w$ term can be ignored. The main task to be performed is then a Fourier Transform. This is nearly always best implemented via an FFT. The next question is how best to deposit the irregularly sampled uv points onto the regular grid used in the FFT. Usually an anti-aliasing filter (AAF) is used to deposit visibility samples onto the grid. This is usually called convolutional gridding and the AAF is called the gridding convolution function (GCF).  Modern practice is to use a prolate spheroidal wavefunction as the AAF.

If the $w$ term is significant, then the GCF can be extended to include the effects of the $w$ term. Numerically this involves multiplying $G(l,m,w)$ with the transform of the AAF and then transforming back to Fourier space. The $w$ projection algorithm uses $\tilde G(u,v,w)$ as a convolutional gridding function.

Since $w$ projection is a convolution in Fourier (data) space, there is an alternate approach in which the data are partitioned in $w$, and $G(l,m,w)$ is applied in image space by multiplying each $w$ plane. We call this $w$ stacking.

The primary beam $A(l,m)$ can be treated in a similar way since it too is a multiplicative term in image space:

\begin{equation}
\label{eqn:aprojrel}
\begin{array}{l}
V(u,v,w)=\tilde A(u,v,w)\otimes V(u,v)\\
\tilde A(u,v,w)=\int\nolimits A\left(l,m\right)\frac{\strut e^{\;j2\pi w(\sqrt{1-l^2-m^2}-1)}}
{\strut\sqrt{1-l^2-m^2}}\;e^{\;j2\pi\left(ul+vm\right)}\;dldm\\
\end{array}
\end{equation}

For obvious reasons, this has acquired the name `$a$ projection' or with the $w$ term, `$aw$ projection' \cite{Bhatnagar2008}. Hybrid versions are also possible in which, for example, the $w$ term is addressed using stacking, and the primary beam via projection (`$a$ projection/$w$ stack'). The exact resources required for $w$ projection are those needed to calculate and cache the GCF. The memory required can be very substantial and indeed prohibitive. Some saving in memory and substantial savings in computation can be realised from the observations that the non-zero part of the convolution is bounded by a cone, and that most values of $w$ are much less than the maximum.

For application of the primary beam using equation \ref{eqn:aprojrel}, the resources can be even larger. The GCF may vary with frequency and/or time, necessitating even frequent recalculation or caching. Nevertheless, application of the primary beam in this way is very valuable.

\item[Snapshots] For a short or snapshot observation made with a coplanar array, neglect of the w-term results in a distorted image coordinate system \cite{CornwellPerley1992,Ord:2010tt}. This coordinate distortion can be corrected in the image plane by interpolation of the image to the correct coordinate system. For that short period of time, the array will be instantaneously coplanar to good accuracy. This means that the $w$ coordinate is related to $(u,v)$ by a simple relationship:

\begin{equation}
\label{eqn:wplane}
w = a u + b v
\end{equation}

The relationship between visibility and sky brightness may be rewritten as a two-dimensional Fourier transform by introducing distorted coordinates $(l',m')$ where:

\begin{equation}
\label{eqn:distortedlm}
\begin{array}{l}
l' = l + a \left(\sqrt{1- l^2 - m^2}-1\right)\\
m' = m + b \left(\sqrt{1- l^2 - m^2}-1\right)
\end{array}
\end{equation}

Parameters $a,b$ may be estimated by linear fitting to the $u,v,w$ coordinates. For a telescope observing at zenith angle $Z$ and parallactic angle $\chi$,  the parameters $a$ and $b$ are given by:

\begin{equation}
\label{eqn:distortedlmab}
\begin{array}{l}
a = \tan{Z} \sin{\chi}\\
b = - \tan{Z} \cos{\chi}
\end{array}
\end{equation}

If performed as the sole means of correcting the non-coplanar baselines effect, the work required for the image plane reprojection can come to dominate. However, if used in conjunction with $w$ projection (or $aw$ projection), an optimum tradeoff between the resources needed for each may be obtained. We discuss such a hybrid algorithm below.

The image reprojection step must be done with high accuracy so as not to misrepresent the model or residual image. For the prediction step this is particularly important since the model may not be diffraction limited as the residual image is. Our code currently allows only bilinear or bicubic interpolation. Lanczos interpolation \cite{Duchon1979} is likely to be much more accurate, even compared to a truncated sinc function.

\item[Facets] The $w$ phase screen can be taken to be constant or linear over small regions of the sky\cite{CornwellPerley1992}. After applying the residual phase term, an image can be constructed by piecing together many different facets. This approach is expensive in terms of floating point operations but has a modest memory footprint\cite{2008ISTSP...2..647C}. It can also be extended to deal with non-isoplanatism.

\item[$w$ snapshots]

A superior algorithm can be obtained by combining $w$ projection and snapshots. $w$ is expressed as a linear plane plus deviations $\Delta w$.

\begin{equation}
\label{eqn:wplaneh}
w = a u + b v + \Delta w
\end{equation}

The current best plane in $u,v,w$ space is chosen by least squares fit. $w$ projection is used to project all visibilities onto this current plane, thus correcting $\Delta w$, and the snapshot imaging is performed and plane fitting is repeated when the deviation from the last plane exceeds a specified tolerance. 

We call this algorithm `$w$ snapshots'.

\end{description}

The optimum solution depends on the context. For ASKAP, $w$ snapshots provides an optimum use of CPU and memory resources. 

\subsection{Scaling with $R_F$}

In this subsection, we derive and illustrate the fundamental scaling laws for the $w$ term. 

The Fresnel number, $R_F$, measures the $w$ term phase error:

\begin{equation}
\label{eqn:fresnel}
R_F=\frac{\Theta_{FOV}^2}{\Theta_{res}}
\end{equation}

This is roughly equal to 
\begin{equation}
\label{eqn:fresnelbd}
R_F=\frac{\lambda B}{D^2}
\end{equation}

where $\lambda$ is the observing wavelength, $B$ is the baselines, and $D$ is the antenna diameter.

The $w$ term can be considered significant when the Fresnel number (roughly the phase error) is comparable to unity. 

\begin{equation}
\label{eqn:fresnellim}
R_F\sim 1
\end{equation}

However, this is too loose a statement for careful evaluations of imaging performance, and we must introduce another parameter into this relationship.

\begin{equation}
\label{eqn:fresneltightlim}
R_F\sim \alpha_{FOV}
\end{equation}

%
%
%
%
%
%

In practice, these numbers may depend on some very important implicit factors.

\begin{description}
\item[Primary beam sidelobes] To reach high dynamic range the synthesized sidelobes of sources in the first, second, or third primary beam sidelobes may need to be subtracted. In that case the field of view $\Theta_{FOV} \gg \lambda/D$.
\item[High accuracy imaging] To represent and subtract sources with high accuracy may require a more stringent limit on the field of view, in which case the number of pixels quoted above is an overestimate. 
\end{description}


In figure 2, we show the behaviour of the $w$ term as a function of resolution and field of view.
\begin{itemize}
\item The green lines show the size of an image for which the $w$ term can just be ignored.
\item The solid blue line shows the maximum field of view for which the $w$ term can be ignored.
\item The dashed blue lines show the Fresnel number $R_F$ as a function of resolution. This measures the strength of the $w$ term phase error. Larger Fresnel numbers require more computation for all algorithms. 
\item All of these numbers are approximate, depending on the definition of field of view, resolution, acceptable phase error, {\em etc.}
\end{itemize}

\begin{center}
\begin{figure}
\label{fig:w}
\caption{Behavior of $w$ term for the different SKA telescopes: $R_F=1$ (blue line), relative processing cost $R_F$ (blue dashed line), required number of image pixels on each axis (green). The red lines show the behaviour with frequency, with low frequency to the left. The field of view for both SKA1\_DISH is diffraction limited and that for SKA1\_SURVEY is set by the size of the phased array feed. For SKA1\_AA\_LOW we show two curves - that for a single beam and that for 480 beams.}
\includegraphics[scale=0.50,angle=0]{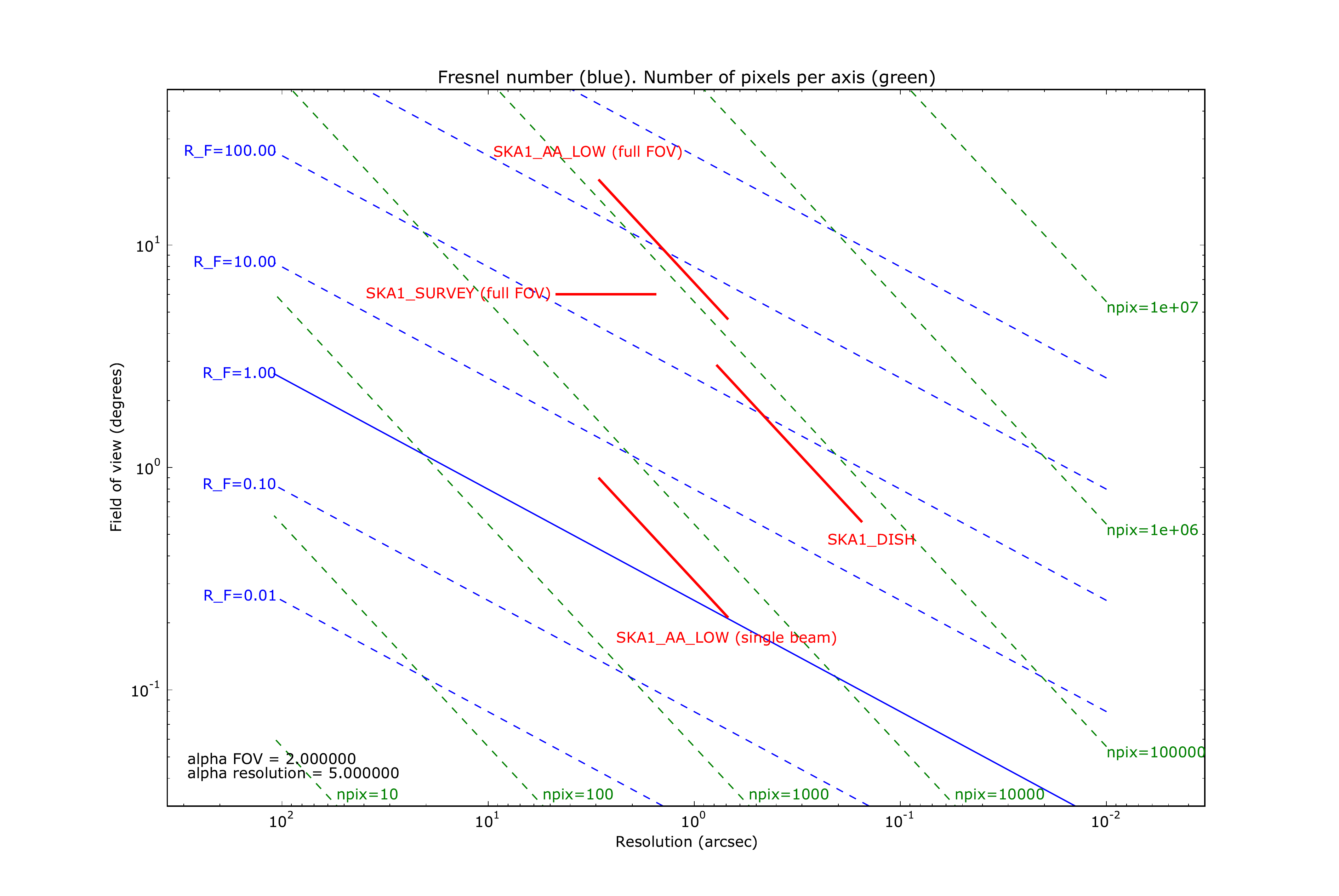}
\end{figure}
\end{center}

To the top right of this diagram, we see observations for which the processing will be inordinately expensive. Imaging in these regions is only possible if the amount of visibility data is sufficiently low.

\section{SCIENTIFIC PERFORMANCE}

The scientific performance can be quantified in multiple ways. Here we concentrate on smearing. All the wide-field algorithms described above are subject to decorrelation losses for sources far from the phase centre. Clearly this can have as deleterious an effect on the science as time and frequency smearing. In this subsection, we investigate the smearing for snapshot imaging and $w$ projection.

\subsection{Snapshot imaging}

For snapshot imaging, we wish to determine the maximum value of $w_{step}$ allowed before a point source drops in intensity by $\Delta A$. 

The shift is a complicated function of parallactic angle and zenith angle. Since we are interested in the order of magnitude of the smearing, we will consider a simplified case. The shifts in position for a source that transits, seen at zenith angle $Z$= 45 degrees, are:

\begin{equation}
\label{eqn:snapshiftl}
\Delta l   = - \Delta \chi a \frac{\Theta^2}{2} =  0
\end{equation}

\begin{equation}
\label{eqn:snapshiftm}
\Delta m = + \Delta \chi b \frac{\Theta^2}{2} = + \Delta \chi \frac{\Theta^2}{2}
\end{equation}

Using a parabolic approximation to the PSF peak and using equations 7 and 9, we find by differentiation that the drop in amplitude is given by:

\begin{equation}
\label{eqn:Aquad}
\Delta A = \frac{1}{2}\left(\frac{\Delta m}{\Theta_{res}}\right)^2
\end{equation}

\begin{equation}
\label{eqn:lquad}
\Delta l = \Theta_{res} \sqrt{2 \Delta A} \\
\end{equation}

Choosing the worst case at the edge of the field:

\begin{equation}
\label{eqn:chiquad}
\Delta \chi = \sqrt{8} \frac{\Theta_{res}}{\Theta_{FOV}^2 } \sqrt{\Delta A}
\end{equation}

Converting a change in parallactic angle to the equivalent increment in w, we find:

\begin{equation}
\label{eqn:wsnapA}
w_{step} = w_{rms} \left(\frac{\sqrt{8\Delta A}}{R_F}\right)
\end{equation}

This is a very stringent limit for high Fresnel numbers and high accuracy. 

We can now calculate the number of snapshots required. Suppose that we observe for hour angle range $h_{obs}$. Then:

\begin{equation}
\label{eqn:nsnap}
N_{step} = h_{obs} \frac{w_{rms}}{w_{step}} =h_{obs} \frac{R_F}{\sqrt{8 \Delta A}}
\end{equation}

The same caveats apply here as for the calculation of optimum $w_{max}$: these equations are approximate only. More accurate values will require simulation.

Snapshot imaging has one substantial flaw. In the duration of the snapshot, the $(u,v,w)$ coverage and hence the best fit plane can change. Hence source positions can be biased and the source structure can be smeared, to a level depending on the duration of the snapshot. The bias can in principle be reduced by fitting to a extended time-chunk of $(u,v,w)$ but this approach would complicate the processing logic. The smearing cannot be easily fixed.

Furthermore, if an array is actually mildly non-coplanar (because of the spherical earth, for example) then there will be uncorrected phase errors, leading to image blurring. This effect will be especially strong at low elevation angles where projection magnifies any non-coplanarity.

\begin{center}
\begin{figure}
\label{fig:positionerrors}
\caption{Position and shape errors incurred by use of snapshot imaging. Top panel shows a point source imaged with various values of $w_{step}$, superimposed on an image of the true source. Bottom shows the same but with $w$ projection added into the processing.}
\begin{center}
\includegraphics[scale=0.65, angle=0]{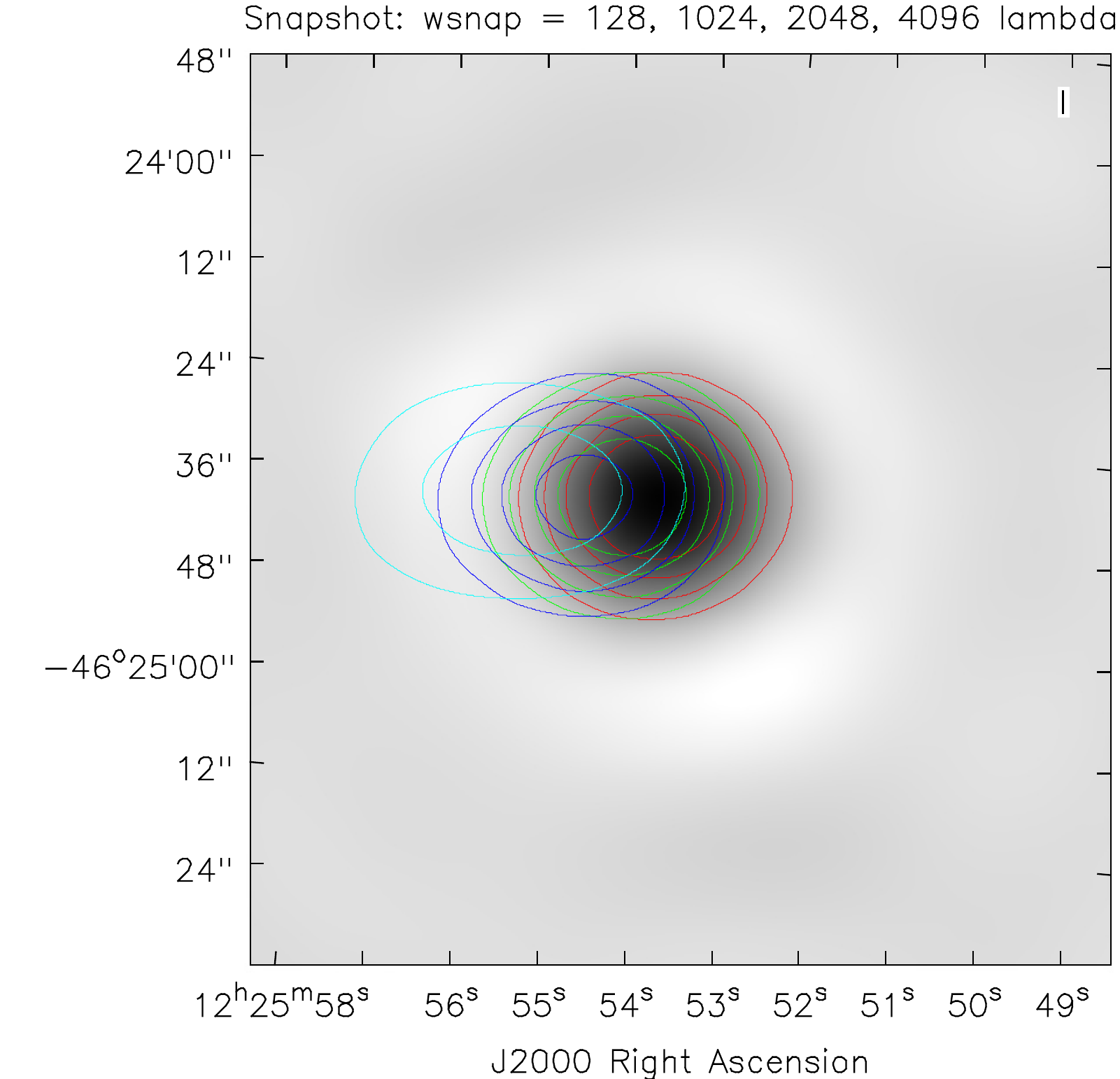}
\includegraphics[scale=0.65, angle=0]{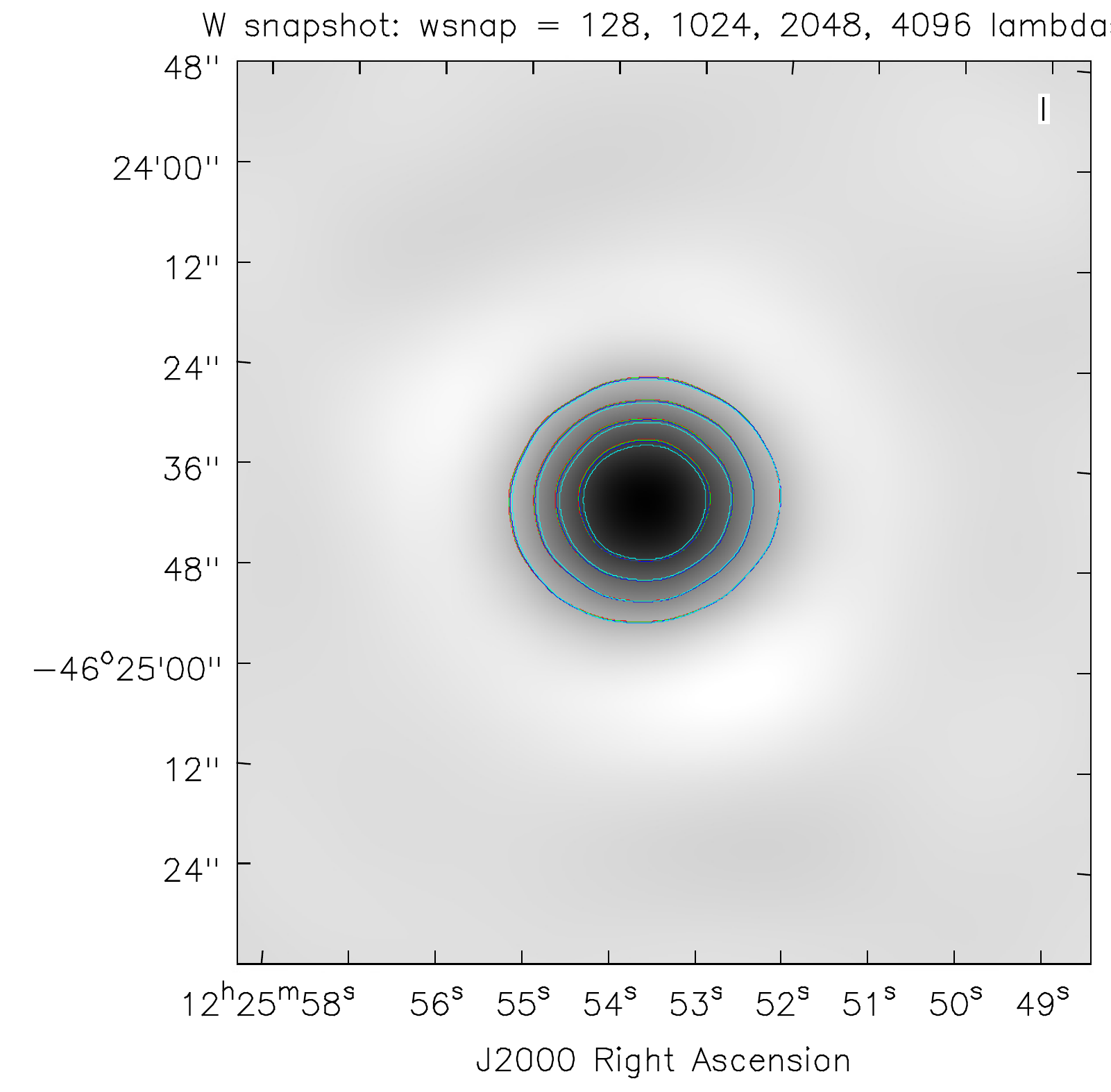}
\end{center}
\end{figure}
\end{center}

\subsection{$w$ projection}

A point source imaged via $w$ projection is subject to decorrelation since the $w$ projection kernel is sampled as discrete locations in u,v,w space and the value for neighbouring grid point points is used in place of the actual value. 
%
%

The convolution function may be implemented via direct calculation as needed or via a pre-calculated lookup table. The accuracy of the lookup table is limited by memory. Hence $w$ projection can be made arbitrarily accurate in a way that the other methods cannot. This means that $w$ projection is very well suited to be used in tandem with another algorithm such as snapshot imaging that suffers significant decorrelation errors. An additional advantage is that the balance between snapshot imaging and $w$ projection can be adjusted to stay within memory and CPU constraints.

The sampling requirements are:

\begin{eqnarray}
\Delta u = \frac{\sqrt{ 2 \Delta A}}{ 2 \pi \theta_{FOV}} \\
\Delta v = \frac{\sqrt{ 2 \Delta A}}{ 2 \pi \theta_{FOV}} \\
\Delta w = \frac{\sqrt{ 2 \Delta A}}{ \pi \theta^2_{FOV}}
\end{eqnarray}

Hence the number of $w$ planes required goes as:

\begin{equation}
N_w = \frac{w_{max} \sin{Z}}{\Delta w}
\end{equation}  

Thus, just as for snapshot imaging, the effect of the $w$ term may be limited by avoiding high zenith angles (low elevations).

\section{COMPUTING PERFORMANCE}

We now turn to investigate the computing performance - specifically the number of operations needed. We will ignore that for the FFT since it nearly always is a minor expense.

\subsection{Processing time for $w$ projection}

The time for $w$ projection goes as:

\begin{equation}
\label{eqn:twprojection}
T_{wprojection} \sim N_{vis} \mu_{wproj} R_F^2
\end{equation}

Note that $\mu_{wproj}$ is for one visibility to one grid point. Typically for gridding a billion visibilities each to one grid cell is $\mu_{wproj}$ ~ 3 - 8 s. There is no dependency on the desired accuracy $\Delta A$ since the accuracy is determined by the precomputed convolution function.

\subsection{Processing time for snapshot imaging}

Let the maximum and rms $w$ baseline be $w_{max}$, $w_{rms}$. When the rms baseline has moved $w_{step}$ then the plane must be transformed, reprojected, and accumulated. Let the range in hour angle be $h_{obs}$ radians. For snapshot imaging to construct an image over an hour angle range $h_{obs}$, the time scales directly as the number of reprojections:

\begin{equation}
\label{eqn:tsnapshot}
T_{snapshot} \sim N_{vis} \mu_{wproj} R_{aa}^2 + N^2_{pixel} \mu_{reproj} h_{obs} \frac{R_F}{\sqrt{8 \Delta A}}
\end{equation}

where $R_{aa}$ measures the size of the anti-aliasing filter in pixels (typically 7 - 9 per axis).

Empirically we find that on a typical desktop, the time to reproject an image of size 8192 by 8192 is about 200s  so, scaling to a million pixels, we find that the cost of reprojecting a million pixels is $\mu_{reproj}\sim 3$s.

The scaling with $R_F$ is only linear, compared to quadratic for $w$ projection, but note that the required accuracy $\Delta A$ is a factor.

\subsection{Processing time for $w$ snapshots}

The calculation for the $w$ projection part of the processing time depends on the rms $w$ after applying the cutoff $w_{step}$. For the moment, we will assume that it is $w_{step}$. Then the time for $w$ snapshots goes as:

\begin{equation}
\label{eqn:htwprojection}
T_{w snapshots} \sim N^2_{pixel} \mu_{reproj} h_{obs} \left(\frac{w_{rms}}{w_{step}}\right)  
+ N_{vis} \mu_{wproj} R_F^2 \left(\frac{w_{step}}{w_{rms} }\right)^2
\end{equation}

Or if $\rho = w_{step}/w_{rms}$: 

\begin{equation}
\label{eqn:htwprojection1}
T_{w snapshots} \sim \frac{N^2_{pixel} \mu_{reproj} h_{obs}} {\rho}  
+ N_{vis} \mu_{wproj} R_F^2 \rho^2
\end{equation}

The optimum value of $\rho$ is:

\begin{equation}
\label{eqn:rhoopt}
\rho_{opt} \sim \left(\frac{N^2_{pixel} \mu_{reproj} h_{obs}} {2 N_{vis} \mu_{wproj} R_F^2}\right)^{\frac{1}{3}}
\end{equation}

And for this value, the processing time is (ignoring terms O(1)):

\begin{equation}
\label{eqn:Topt}
T_{w snapshots,opt} \sim \left( \left(N^2_{pixel} \mu_{reproj} h_{obs}\right)^2 N_{vis} \mu_{wproj} R_F^2\right)^{\frac{1}{3}}
\end{equation}

Thus $w$ snapshots has substantially better scaling with $R_F$ than either $w$ projection or snapshots. The improvements in absolute performance are:

\begin{equation}
\label{eqn:Toptratio}
\frac{T_{w snapshots,opt}}{T_{wprojection}} \sim \left(\frac{N^2_{pixel} \mu_{reproj} h_{obs}}{ N_{vis} \mu_{wproj} R_F^2}\right)^{\frac{2}{3}}
\end{equation}

\begin{equation}
\label{eqn:Toptratio}
\frac{T_{w snapshots,opt}}{T_{snapshot}} \sim \frac{\sqrt{8 \Delta A}}{R_F} \left(
\frac{ N_{vis} \mu_{wproj} R_F^2}{N^2_{pixel} \mu_{reproj} h_{obs}} \right)^{\frac{1}{3}}
\end{equation}


We can illustrate these results by simulations. In figure 4, we show results from a simulated long observation with the SKA1\_AA\_LOW, for baselines up to 40000$\lambda$. The total run time shown here is the sum of the times for regridding images, degridding visibilities, and constructing the $w$ projection convolution functions. The curves show different integration times: 1s (red), 3s (green), 10s (blue). For shorter integration times, the need to grid more visibility points shifts the optimum transition point lower, thus favouring snapshot imaging over $w$ projection. For the 1s integration case, the improvement of $w$ snapshots over pure $w$ projection is substantial - just under a factor of 3. For all integration times, small values of the transition point (less than 10000 in this example) are disfavoured. 

Although the timing ratios can be expected to be qualitatively correct, in practice, it may be preferable to search numerically for the optimum value of $w_{step}$, perhaps as an autotuning procedure.

\begin{center}
\begin{figure}
\label{fig:simonly}
\caption{Execution time as function of transition from snapshot imaging to $w$ projection. The curves show different integration times: 1s (red), 3s (green), 10s (blue). For shorter integration times, the need to grid more visibility points shifts the optimum transition point lower, thus favouring snapshot imaging over $w$ projection. }
\begin{center}
\includegraphics[scale=0.5, angle=0]{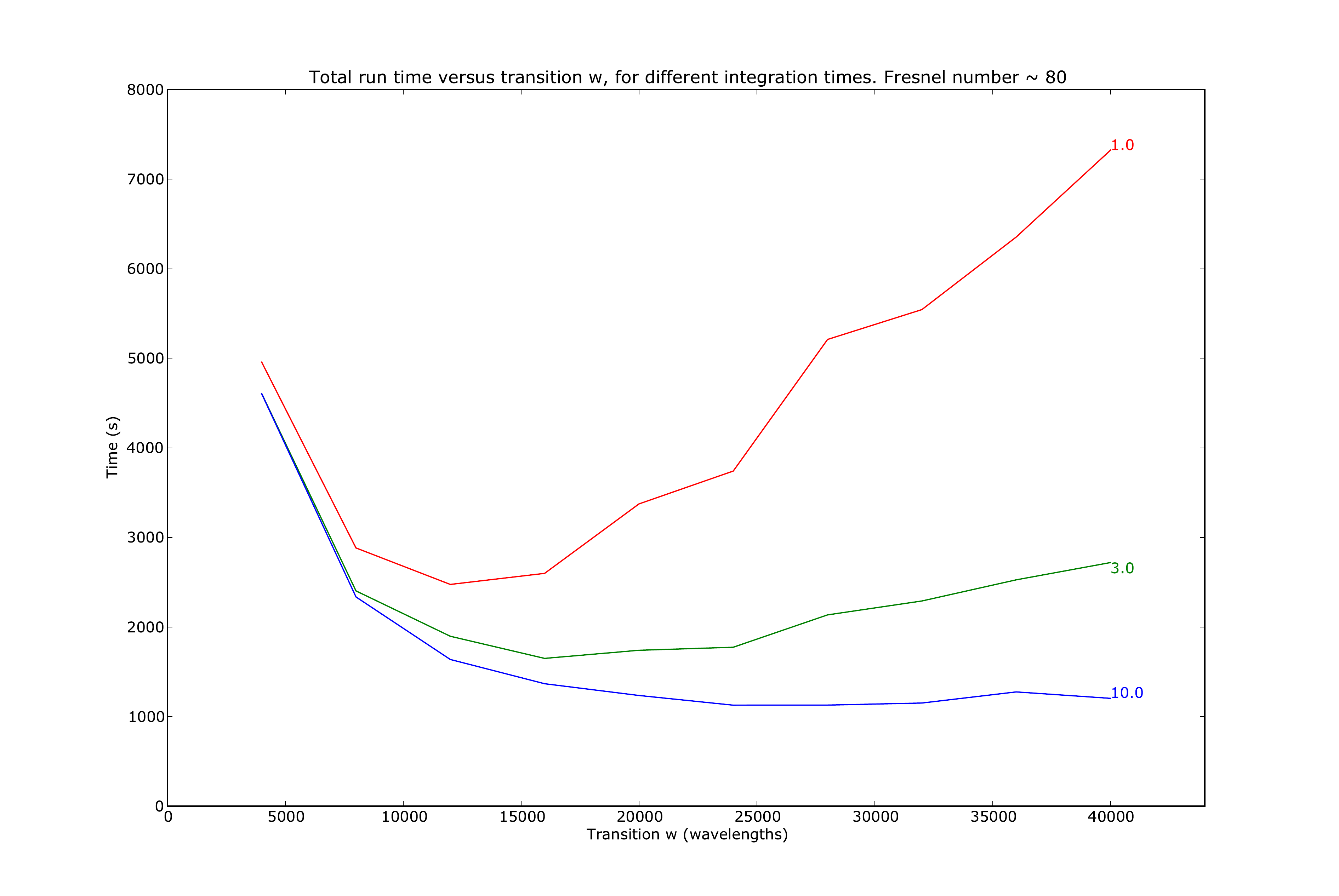}
\end{center}
\end{figure}
\end{center}

\begin{table}
\label{tab:scaling}
\caption{Scaling of processing time with Fresnel number $R_F$}
\begin{center}
\large
\begin{tabular}{|c|c|}
\hline
Algorithm & Scaling of time \\
\hline
$w$ projection & $R^2_F$ \\
snapshots & $R_F$ \\
$w$ snapshots & $ R^{2/3}_F $ \\
\hline
\end{tabular}
\end{center}
\end{table}

\section{DISCUSSION}
\label{sec:discussion}

Table 1 summarises the scaling with $R_F$ for the three algorithms. Our new algorithm, $w$ snapshots, has considerably better scaling than $w$ projection, and has better scaling than snapshots, while avoiding the shape distortions and biases of the latter. If we take the number of visibilities and the total number of pixels to be the same, $h_{obs}\sim 1$, and also take $\mu_{reproj} \sim \mu_{wproj}$, then $w$ snapshots is faster than $w$ projection by a factor $R^{4/3}_F$.

Furthermore, $w$ snapshots is superior to snapshot imaging for few visibilities, many pixels, and high desired accuracy, and is superior to $w$ projection for few pixels, many visibilities. 

The processing time (equation \ref{eqn:Topt}) contains terms related to the observation $N^2_{pixel}, h_{obs}, N_{vis}, R_F$ and terms related to the computer performance $\mu_{reproj}, \mu_{wproj}$. The latter two terms can be optimised. We have expended significant effort\cite{Humphreys2011} on optimising the gridding cost $\mu_{wproj}$ but no effort on $\mu_{reproj}$ so there may be significant improvements yet to be had.

Referring back to figure 2, we can see that these improvements in scaling will have the effect of decreasing the computational cost for SKA imaging substantially, an order of magnitude or more for SKA1\_DISH, for example. Verifying this prediction will require improvement in the ability of our software to handle very large images ($10^9-10^{10}$ pixels), and is deferred to a subsequent paper.

\section*{ACKNOWLEDGEMENTS}
\label{sec:acknowledgements}

We thank all our colleagues in the ASKAP Computing group for their contributions to this work. 

The computational work has been made possible by a grant from the CSIRO share of the National Computational Infrastructure National Facility. In addition, this work was supported by the iVEC@Murdoch supercomputer, Epic.

\bibliographystyle{spiebib}
\bibliography{hybridw}

\end{document}